\documentclass[pra,twocolumn,floatfix,showpacs,superscriptaddress]{revtex4}

\usepackage{amsmath}
\usepackage{amsfonts}
\usepackage{graphicx}

\DeclareMathSymbol{\N}{\mathalpha}{AMSb}{'116}
\DeclareMathSymbol{\R}{\mathalpha}{AMSb}{'122}

\begin{document}

\title{Effective disentanglement of measured system and measurement 
apparatus}
\author{S. Camalet}
\affiliation{Laboratoire de Physique Th\'eorique de la Mati\`ere Condens\'ee, 
UMR 7600, Universit\'e Pierre et Marie Curie, Jussieu, Paris-75005, France}
\date{Received: date / Revised version: date }
\begin{abstract}
We consider a multi-level system coupled to a bosonic measurement 
apparatus. We derive exact expressions for the time-dependent expectation 
values of a large class of physically relevant observables that depend on 
degrees of freedom of both sytems. We find that, for this class, though 
the two systems become entangled as a result of their interaction, 
they appear classically correlated for long enough times. The unique 
corresponding separable state is determined explicitly. To better 
understand the physical parameters that control the time scale of 
this effective disentanglement process, we study a one-dimensional 
measurement apparatus.
\end{abstract} 


\maketitle

\section{Introduction}

As is well known, interactions between quantum systems tend to increase 
their entanglement. Quantum correlations between physical systems 
should then be omnipresent. A first obstacle to detecting them is that 
real systems are inevitably influenced by surrounding degrees of freedom. 
The importance of the role played by the environment is substantiated by 
the fact that two systems cannot remain maximally entangled while they 
get entangled with a third system \cite{CKW}. And indeed, it has been 
shown, for both free particles \cite{DH} and two-level systems \cite{JJ} , 
that two non-interacting open systems, initially prepared in an entangled 
state, evolve into a classically correlated state. However, when interactions 
between the two systems are taken into account, the situation is not that 
clear. Revivals of entanglement and even long-time entanglement have 
been obtained  \cite{RR,FT,WBPS,KAM}. Moreover, even if there is no 
direct interaction, entanglement can be induced by environment-mediated 
interactions \cite{RR,BFP}. The influence of the environment may thus not 
fully explain why quantum correlations are so imperceptible. 

In the above-cited works, the correlations between the two systems 
considered are studied using their full bipartite quantum state. Such 
complete knowledge is unattainable when the systems of interest consist 
of a large number of degrees of freedom. In general, the accessible 
information on the state of the compound system under study consists of 
a finite set of expectation values. Such limited data can be compatible with 
a classically correlated state whereas the actual bipartite state is entangled 
\cite{HHH,AP}. Interaction-induced quantum correlations may thus be 
practically undetectable, even in the case of negligible influence of 
the environment, if one or both of the two coupled systems is large enough.

A prominent example of such a situation is provided by the dynamical 
approach to the measurement process. The reduced state of a system 
${\cal S}$ suitably coupled to a larger one ${\cal M}$, evolves into 
a statistical mixture of pure states determined by the interaction between 
${\cal S}$ and ${\cal M}$, with weights given by Born rule \cite{Z,E1,E2}. 
This decoherence is directly related to the development of entanglement 
between ${\cal S}$ and ${\cal M}$. However, as mentioned above, 
quantum correlations between these two systems may be essentially 
indiscernible.

In this paper, we address this issue by considering a measurement 
apparatus ${\cal M}$ that consists of harmonic oscillators. The resulting 
model is simple enough to allow the derivation, without any 
approximation, not only of the reduced dynamics of ${\cal S}$, which is 
the usual focus of decoherence studies \cite{LCDFGZ,QDS,SADH,EPJB}, 
but also of the temporal evolution of correlations between ${\cal S}$ and 
${\cal M}$ induced by their mutual interaction. The paper is organized as 
follows. The model we consider and some of its features are presented in 
the next section. In Sec. \ref{sec:OI}, physically relevant observables of 
the complete system ${\cal S}+{\cal M}$ are introduced and exact 
expressions for their time-dependent expectation values are derived. 
We will see that, in parallel to the decoherence of ${\cal S}$, quantum 
correlations between ${\cal S}$ and ${\cal M}$ decay with time. This result 
is obtained for a generic measured system ${\cal S}$ and under the only 
assumption that the measurement apparatus ${\cal M}$ is bosonic. 
In order to better understand what determines the time scale of this process, 
we study in some detail the special case of a two-level system ${\cal S}$ 
coupled to a one-dimensional free field system ${\cal M}$ in Sec. 
\ref{sec:1DMA}. Finally, in the last section, we summarize our results 
and mention some questions raised by our work.
   
\section{Measurement Model}\label{sec:MM}

The complete system consisting of the measured system ${\cal S}$ and 
measurement apparatus ${\cal M}$ is described by the Hamiltonian
\begin{multline}
H = \sum_\ell E_\ell |\ell \rangle \langle \ell | + 
\sum_q \omega_q a^{\dag}_q a^{\phantom{\dag}}_q \\
+ \sum_{\ell,q} |\ell \rangle \langle \ell | \otimes
\left[ \lambda_{\ell q} a^{\dag}_q + \lambda_{\ell q}^* 
a^{\phantom{\dag}}_q \right] 
\label{H}
\end{multline}
where the annihilation operators $a_q$ satisfy the bosonic commutation 
relations $[a_q,a_{q'}]=0$ and 
$[a^{\phantom{\dag}}_q,a^{\dag}_{q'}]=\delta_{qq'}$, and $E_\ell$ and 
$|\ell \rangle$ are the eigenenergies and eigenstates of ${\cal S}$. 
We define for further use the Hamiltonian 
$H_0=\sum_q \omega_q a^{\dag}_q a^{\phantom{\dag}}_q$ 
which characterizes ${\cal M}$ in the absence of interaction with 
${\cal S}$ and the measurement apparatus Hamiltonians 
\begin{equation}
H_\ell=H_0 + \sum_{q} \left[ \lambda_{\ell q} a^{\dag}_q 
+ \lambda_{\ell q}^* a^{\phantom{\dag}}_q \right]  .
\label{Hell}
\end{equation} 
We assume that, initially, ${\cal S}$ and ${\cal M}$ are  uncorrelated 
and ${\cal M}$ is in thermal equilibrium with temperature $T$, i.e., 
the system ${\cal S}+{\cal M}$ is, at time $t=0$, in the state
\begin{equation}
\Omega = \sum_{\ell , \ell'} \rho_{\ell \ell'} |\ell \rangle\langle \ell'|
\otimes Z^{-1} e^{-H_0/T} \label{Omega}
\end{equation}
where $Z= \mathrm{Tr} \exp(-H_0/T)$ and 
$\sum_{\ell,\ell'} \rho_{\ell \ell'} |\ell \rangle\langle \ell'|$ is any state 
of ${\cal S}$. Throughout this paper, we use units in which 
$\hbar=k_B=1$. 

\subsection{Interaction-induced entanglement}\label{subsec:Iie}

If ${\cal S}$ is initially in one of its eigenstates $|\ell \rangle$, ${\cal S}$ 
and ${\cal M}$ remain uncorrelated and the state of ${\cal S}$ stays 
equal to $|\ell \rangle$ as required for the measurement of an observable 
with eigenstates $|\ell \rangle$. But this is a very particular case. 
In general, ${\cal S}$ and ${\cal M}$ become entangled under the action 
of the Hamiltonian \eqref{H}. This Hamiltonian has the generic property 
$[H_\ell,H_0] \ne 0$, and hence, contrary to measurement models 
such that these commutators vanish, the thermal statistical average 
in \eqref{Omega} is not essential for the decoherence of ${\cal S}$ 
which persists at zero temperature \cite{E1,E2}. As mentioned 
in the introduction, the fundamental origin of this decoherence is 
the evolution of the entanglement between  ${\cal S}$ and ${\cal M}$. 
For example, at $T=0$ and for a two-level system ${\cal S}$ initially in 
the pure state $2^{-1/2}(|1 \rangle +|2 \rangle)$, the (pure) state of 
${\cal S}+{\cal M}$ at time $t$ reads in Schmidt form 
\begin{equation}
|\Psi (t) \rangle = \sum_{\eta=\pm} [1+\eta F_{12}(t)]^{1/2}
\big( |1 \rangle +\eta |2 \rangle \big) |\psi_\eta (t) \rangle /2 \nonumber
\end{equation}
where $|\psi_\pm \rangle$ are states of ${\cal M}$ obeying 
$\langle \psi_\eta |\psi_{\eta'} \rangle=\delta_{\eta \eta'}$ and $F_{12}$ is 
directly related to the decoherence of ${\cal S}$. We will see below that 
$F_{12}$ decays from $1$ to $0$ as time goes on. Thus, the above state 
$|\Psi \rangle$ evolves from a product state to a maximally entangled one 
\cite{fn0}. 

\subsection{Complete system expectation values}

In the general case, the state of ${\cal S}+{\cal M}$ at arbitrary time $t$ 
is mixed and entangled. We are interested in the resulting expectation 
values $\langle O \rangle (t)=\mathrm{Tr} [\exp(-itH) \Omega \exp(itH) O ]$ 
of observables $O$ of the complete system ${\cal S}+{\cal M}$. 
We expand them as 
\begin{equation}
O = \sum_{\ell,\ell'}   
|\ell \rangle\langle \ell'| \otimes O_{\ell\ell'}  
\end{equation}
where $O_{\ell\ell'}$ are operators acting in the Hilbert space of ${\cal M}$, 
that obey $O_{\ell'\ell}=O_{\ell\ell'}^\dag$. With these notations, 
their expectation values can be written as
\begin{multline}
\langle O \rangle (t) = \sum_{\ell}  \rho_{\ell \ell} 
\big\langle e^{i t H_\ell} O_{\ell \ell} e^{-i t H_\ell} \big\rangle_{\cal M}  
\label{Ot}  \\ 
+ 2 \mathrm{Re} \sum_{\ell<\ell'}  \rho_{\ell' \ell} 
e^{it(E_\ell-E_{\ell'})} \big\langle e^{i t H_\ell} O_{\ell \ell'} 
e^{-i t H_{\ell'}}  \big\rangle_{\cal M}  
\end{multline} 
where $\langle \ldots \rangle_{\cal M}=\mathrm{Tr}(\exp(-H_0/T) \ldots )/Z$, 
since $H=\sum_\ell |\ell \rangle \langle \ell | (E_\ell + H_\ell)$. If $O$ is 
an observable of ${\cal S}$ alone, the $O_{\ell \ell'}$ are simple numbers 
and the first term of \eqref{Ot} is constant. In contrast, the second term of 
this expression can vanish at long times. The reduced state of ${\cal S}$ is 
then a statistical mixture of the states $|\ell \rangle$ with weights 
$\rho_{\ell \ell}$ as expected after an unread measurement. In other words, 
${\cal S}$ decoheres. We show in the following that the second term of 
\eqref{Ot} can also vanish asymptotically for true operators $O_{\ell\ell'}$. 
In this case, although ${\cal S}$ and ${\cal M}$ get entangled under 
the action of \eqref{H}, the expectation value $\langle O \rangle (t)$ 
becomes identical to that of the separable state 
\begin{equation}
\Omega_\mathrm{eff} (t) = \sum_{\ell}  \rho_{\ell\ell}
|\ell \rangle \langle \ell| \otimes e^{-i t H_\ell} 
Z^{-1} e^{-H_0/T}  e^{i t H_\ell} 
\label{Oefft}
\end{equation}
which is a statistical mixture of the product states 
$ |\ell \rangle  \exp(-it H_\ell) |\{ n_q \} \rangle$ where $| \{ n_q \} \rangle$ 
are the eigenstates of $H_0$. The correlations between ${\cal S}$ and 
${\cal M}$ described by such a state are of classical nature \cite{W}. 
Remark that for an observable $O$ of ${\cal M}$ alone, 
i.e., $O_{\ell\ell'}=O \delta_{\ell\ell'}$, there is no difference between 
\eqref{Oefft} and the actual state of ${\cal S}+{\cal M}$.

\section{Observables of interest}\label{sec:OI}
 
Many physical systems can be modeled by the Hamiltonian \eqref{H}. 
The corresponding bosonic field can be, for instance, the electromagnetic 
field \cite{CDG}, the atomic displacement field of a crystal \cite{QDS} or 
the charge distribution of an LC transmission line \cite{EPL}. We consider 
observables $O$ which are functions of operators of the form 
\begin{equation}
\Pi_\alpha = \sum_q \left[ \mu_{\alpha q} a^{\dag}_q 
+ \mu_{\alpha q}^* a^{\phantom{\dag}}_q \right] . \label{Pi}
\end{equation}
Such linear combinations of creation and annihilation operators can be 
interpreted as local components of the bosonic field described by $H_0$. 

\subsection{Generating functions}

In order to obtain the contribution of any product 
$\prod_\alpha (\Pi_\alpha)^{n_\alpha}$ where $n_\alpha \in \N$, to 
the expectation value \eqref{Ot}, we define the generating functions  
\begin{equation}
K_{\ell\ell'}(t;\{X_\alpha\})=\Big\langle e^{i t H_\ell} 
\prod_\alpha \exp( i X_\alpha \Pi_\alpha ) e^{-itH_{\ell'}} 
\Big\rangle_{\cal M} \label{Kdef}
\end{equation}
where the $X_\alpha$ are real numbers. These averages can be evaluated 
by noting that the Hamiltonian \eqref{Hell} and $H_0$ are related by 
a unitary transformation :
\begin{equation}
H_\ell = U_\ell H_0 U^\dag_\ell -\sum_q 
\big| \lambda_{\ell q} \big|^2/\omega_q^2 
\label{U}
\end{equation}
where $U_\ell=\prod_q \exp[ (\lambda_{\ell q}^* a^{\phantom{\dag}}_q
-\lambda_{\ell q} a^{\dag}_q)/\omega_q]$, and by using 
$\langle \exp(z a_q - z^* a^{\dag}_q ) \rangle_{\cal M}
=\exp[-|z|^2/2\tanh(\omega_q/2T)]$ where $z$ is any complex number. 
For $\ell=\ell'$, the calculation is straigthforward and gives
\begin{equation}
K_{\ell\ell}(t;\{X_\alpha\})=
\exp \Big( 2 i \sum_\alpha X_\alpha A_\alpha^{(\ell)}(t) 
-\sum_{\alpha \le \alpha'} X_\alpha X_{\alpha'} C_{\alpha\alpha'}  
\Big) \label{Kdiag}
\end{equation}
where 
$C_{\alpha\alpha'}=\langle \Pi_\alpha \Pi_{\alpha'} \rangle_{\cal M} $ 
(for $\alpha \ne \alpha'$) are the correlations of the observables 
\eqref{Pi} at thermal equilibrium, 
$C_{\alpha\alpha}=\langle \Pi_\alpha^2 \rangle_{\cal M}/2 $ and
\begin{equation}
A_\alpha^{(\ell)}(t) =  \mathrm{Re} \int_0^\infty d\omega   
{\cal G}_\alpha^{(\ell)}(\omega) \left(e^{i\omega t}-1 \right)/\omega . 
\label{A}
\end{equation}
Details are given in Appendix \ref{appsec:Dgf}. In the above expression, 
we have introduced the frequency function 
$ {\cal G}_\alpha^{(\ell)}(\omega)=\sum_q \mu_{\alpha q} \lambda_{\ell q}^* 
\delta(\omega-\omega_q)$. 
For a large system ${\cal M}$, it can be regarded as a continuous function. 
For $\ell \ne \ell'$, \eqref{Kdiag} generalises to 
\begin{multline}
K_{\ell\ell'}(t;\{X_\alpha\})= F_{\ell\ell'}(t)
\exp \Big( -\sum_{\alpha \le \alpha'} X_\alpha X_{\alpha'} C_{\alpha\alpha'} \\
+ \sum_\alpha X_\alpha \big[ iA_\alpha^{(\ell)}(t)+ iA_\alpha^{(\ell')}(t) 
- B_\alpha^{(\ell)}(t)+B_\alpha^{(\ell')}(t) \big] \Big) \label{K}
\end{multline} 
where
\begin{eqnarray} 
B_\alpha^{(\ell)}(t) &=& \mathrm{Im} \int_0^\infty d\omega 
\frac{{\cal G}_\alpha^{(\ell)}(\omega) (e^{i\omega t}-1)}
{ \tanh(\omega/2T)\omega}  \label{B} \\
|F_{\ell\ell'}(t)| &=& \exp \left[ - 2 \int_0^\infty d\omega 
\frac{{\cal J}_{\ell\ell'}(\omega) 
\sin^2(\omega t/2)}{\tanh(\omega/2T) \omega^2} \right] \label{F} 
\end{eqnarray}
with ${\cal J}_{\ell\ell'}(\omega)=\sum_q |\lambda_{\ell q}-\lambda_{\ell' q}|^2 
\delta(\omega-\omega_q)$. The derivation of \eqref{K} and the phase of 
$F_{\ell\ell'}$ can be found in Appendix \ref{appsec:Dgf}. Remark that 
the functions \eqref{A}, \eqref{B} and \eqref{F} are finite only if 
${\cal J}_{\ell\ell'}$, $\mathrm{Re} {\cal G}_\alpha^{(\ell)}$ and 
$\omega \mathrm{Im} {\cal G}_\alpha^{(\ell)}$ go to zero for 
$\omega \rightarrow 0$. We also observe that \eqref{A} and \eqref{B} can 
be written in terms of the thermal time-dependent correlation function of 
the observables $\Pi_\alpha$ and 
$\Pi_\ell=\sum_{q} [ \lambda_{\ell q} a^{\dag}_q 
+ \lambda_{\ell q}^* a^{\phantom{\dag}}_q ] $ as
\begin{equation}
B_\alpha^{(\ell)}(t)-iA_\alpha^{(\ell)}(t) =  \int_0^t dt' 
\langle  \Pi_\ell \Pi_{\alpha} (t') \rangle_{\cal M} \label{PilPia}
\end{equation}
where $\Pi_{\alpha} (t)=\exp(itH_0)\Pi_{\alpha}\exp(-itH_0)$.

\subsection{Decoherence}

For an observable $O_{\cal S}$ of ${\cal S}$ alone, the expectation 
value \eqref{Ot} simplifies to
\begin{equation}
\langle O_{\cal S} \rangle (t) = \sum_{\ell}  \rho_{\ell \ell} O_{\ell \ell} 
+ 2 \mathrm{Re} \sum_{\ell<\ell'}  \rho_{\ell' \ell} e^{it(E_\ell-E_{\ell'})} 
O_{\ell \ell'} F_{\ell \ell'}(t) \nonumber
\end{equation} 
where here the $O_{\ell \ell'}$ are simple numbers. The long time behavior 
of this average is governed by the low frequency behaviors of the spectral 
densities ${\cal J}_{\ell\ell'}$. We assume as usual that, for small $\omega$, 
${\cal J}_{\ell\ell'}(\omega) \sim \omega^s$ where $s>0$ \cite{LCDFGZ}. 
For $s < 2$, $\ln |F_{\ell\ell'}|$ diverges as $t^{2-s}$ at long times, whereas, 
for $s>2$, $F_{\ell\ell'}$ reaches a finite value in this limit \cite{fn1}. 
Consequently, the second term of the above expression vanishes 
asymptotically, and ${\cal S}$ decoheres, if all the spectral densities 
${\cal J}_{\ell\ell'}$ approach zero slowly enough as $\omega \rightarrow 0$. 

\subsection{Effective disentanglement}

Any average 
$\langle \exp(it H_\ell ) \prod_\alpha (\Pi_\alpha)^{n_\alpha} 
\exp(-itH_{\ell'}) \rangle_{\cal M}$ can be obtained by expanding 
the expressions \eqref{Kdef} and \eqref{K} in powers of $X_\alpha$. All 
these expectation values are of the form $F_{\ell\ell'}(t) G(t)$ where $G$ is 
a function of time. As a consequence of the low-frequency behaviors of the 
${\cal G}_\alpha^{(\ell)}$ discussed above, $G$ diverges at most 
algebraically in the long-time limit. Thus, for observables $O$ which can be 
written in terms of finite products $\prod_\alpha (\Pi_\alpha)^{n_\alpha}$, 
the second term of \eqref{Ot} decays with time when ${\cal S}$ decoheres 
\cite{fn1}. 

The conclusion is less clear if, in the series expansion of $O$ in terms of 
$\Pi_\alpha$, the sum over $n_\alpha$ runs to infinity. An interesting 
example of this kind is the joint probability of finding, at time $t$, ${\cal S}$ 
in a given state $| u \rangle = \sum_\ell u_\ell | \ell \rangle$ and a field 
component $\Pi_1$ between $p$ and $p+dp$. This probability reads 
\begin{multline}
\Big\langle | u \rangle\langle u | \otimes \delta(\Pi_1-p) \Big\rangle (t) = 
\frac{1}{2\pi} \sum_{\ell,\ell'} u_\ell u_{\ell'}^* \rho_{\ell' \ell} 
e^{it(E_\ell-E_{\ell'})} \\ \times \int dx e^{-ipx} K_{\ell\ell'} (t;x) .
\end{multline} 
Since $K_{\ell\ell'}$ is Gaussian in $x$, the above Fourier transform is 
readily evaluated and we find
\begin{multline}
\Big\langle | u \rangle\langle u | \otimes \delta(\Pi_1-p) \Big\rangle (t) =
\sum_{\ell}  \frac{\rho_{\ell\ell} |u_\ell|^2}{\sqrt{\pi} \Delta} 
e^{-[{\bar p}-Q_{\ell\ell} (t)]^2} \\
+\sum_{\ell<\ell'}   
\frac{2{\tilde F}_{\ell\ell'}(t)}{\sqrt{\pi} \Delta} e^{-[{\bar p}-Q_{\ell\ell'} (t)]^2} 
\mathrm{Re} \left( u_\ell u_{\ell'}^* \rho_{\ell' \ell} e^{it(E_\ell-E_{\ell'})} \right. 
\\ \left. \times \exp \Big[ 2i\big[ B_1^{(\ell)}(t)- B_1^{(\ell')}(t) \big] 
\big[ {\bar p}-Q_{\ell\ell'} (t)\big]/\Delta \Big] \right) \label{prob}
\end{multline} 
where $\Delta= \sqrt{2 \langle \Pi_1^2 \rangle_{\cal M}}$, 
${\bar p}=p/\Delta$, $Q_{\ell\ell'}  = [A_1^{(\ell)}+A_1^{(\ell')}]/\Delta$ and 
${\tilde F}_{\ell\ell'}=F_{\ell\ell'}\exp([B_1^{(\ell)}- B_1^{(\ell')}]^2/\Delta^2)$. 
We have seen above that the decoherence of ${\cal S}$ is ensured by 
the vanishing of $F_{\ell\ell'}$ in the limit $t \rightarrow \infty$ but 
the long-time behavior of ${\tilde F}_{\ell\ell'}$ depends also on that 
of $B_1^{(\ell)}(t)$ and the general expression \eqref{B} does not exclude 
the possibility that these functions diverge as $t \rightarrow \infty$. 
However, \eqref{PilPia} shows that if the correlation 
$\langle \Pi_\ell \Pi_1 (t) \rangle_{\cal M}$ vanishes fast enough at 
infinity then $B_1^{(\ell)}(t)$ does not diverge and hence the quantum 
interference part of \eqref{prob} disappears with time. A specific system 
${\cal M}$ is studied in the following. 

\subsection{Characteristic time scale}

We now address the issue of the characteristic time scale of the quantum 
interference term of \eqref{Ot}. First, it is clear from the above discussion 
that, for finite products $\prod_\alpha (\Pi_\alpha)^{n_\alpha}$, 
the long-time behavior of this term is essentially determined by the factor 
\eqref{F} and hence that the corresponding effective disentanglement time 
scale is the decoherence time of ${\cal S}$. This is not the case for all 
observables $O$ and the time required for the second term of \eqref{Ot} 
to vanish depends strongly on the observable considered. For example, for 
\begin{equation}
O^{(12)}(t_0)=e^{i(E_2-E_1)t_0} |1 \rangle\langle 2 | \otimes 
e^{-iH_1 t_0} e^{iH_2 t_0}  
+ \mathrm{h.c.}, \label{O12}
\end{equation}
the expectation value 
$\langle O^{(12)} (t_0) \rangle (t)=2\mathrm{Re} \rho_{21} 
\exp[i(E_1-E_2)(t-t_0)] F_{12}(t-t_0)$ is finite at $t=t_0$ and goes to zero at 
infinite time. Therefore, for any given time $t_0$, there exist observables for 
which the second term of \eqref{Ot} is important at $t=t_0$ but eventually 
vanishes for longer times. In other words, effective disentanglement cannot, 
strictly speaking, be characterized by a unique time scale. Interestingly, 
$O^{(12)}(t_0)$ belongs to the class of observables discussed above. It can 
be written in terms of a field operator of the form \eqref{Pi} since 
$\exp(-iH_1 t_0) \exp(iH_2 t_0)=\exp[i\varphi_{12}(-t_0)+i\Xi(t_0)]$ where
\begin{equation}
\Xi (t_0) = i \sum_q (\lambda_{1 q}-\lambda_{2 q}) 
( 1-e^{-i\omega_q t_0} ) a^\dag_q / \omega_q + \mathrm{h.c.} . \label{Xi}
\end{equation}
The phase $\varphi_{\ell\ell'}$ is given in Appendix \ref{appsec:Dgf}.

\section{One-dimensional measurement apparatus}\label{sec:1DMA}

As a simple example of system ${\cal S}+{\cal M}$, let us consider 
a two-level system ${\cal S}$ coupled to a one-dimensional measuring 
device ${\cal M}$ described by the Hamiltonian 
\begin{equation}
H = \frac{1}{2} \int dx \left[ \Pi(x)^2 
+ c^2 (\partial_x \phi)^2 \right]
+ g \sigma_z \int dx h(x) \Pi(x) \label{H1D}
\end{equation}
where the fields $\Pi$ and $\phi$ are canonically conjugate to each other, 
i.e., $[\phi(x),\Pi(x')]=i\delta(x-x')$, $c$ is the field propagation speed, 
$g$ characterizes the coupling strength between ${\cal S}$ and ${\cal M}$, 
and $\sigma_z=|1\rangle\langle 1|-|2\rangle\langle 2|$. The even test 
function $h(x)$ is maximum at $x=0$ and vanishes for $|x| \gg a$. 
The fields $\Pi$ and $\partial_x \phi$ can be interpreted, for example, as 
the electric and magnetic components of a one-dimensional cavity 
electromagnetic field \cite{CDG}, or as the charge and current distributions 
of an LC transmission line \cite{EPL}. The measurement apparatus 
${\cal M}$ is assumed to be initially in its ground state, i.e., $T=0$.

\subsection{Local observables}\label{subsec:Lo}

\begin{figure}
\centering \includegraphics[width=0.45\textwidth]{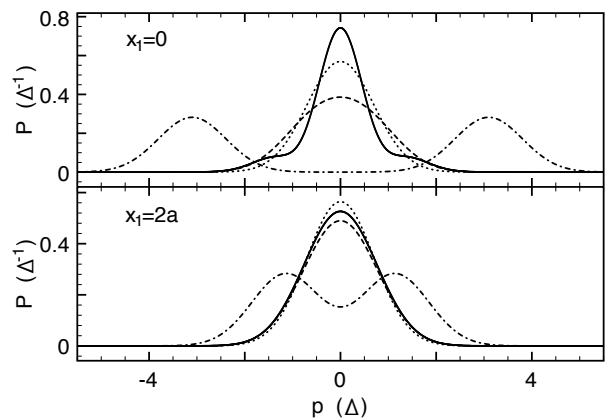}
\caption{\label{fig:disp} Conditional probability $P$ of finding $\Pi_1=p$ 
at time $t$, given ${\cal S}$ is found in its initial state 
$(| 1 \rangle+ | 2 \rangle)/\sqrt{2}$ at the same time, for $t=t_1$ (solid lines) 
and $t=t_2>t_1$ (dash-dotted lines). The dashed lines correspond to 
the separable part of $P$ at $t=t_1$. This contribution is indistinguishable 
from the complete distribution at $t=t_2$. The dotted lines are the initial 
thermal Gaussian distribution. For $x_1=0$, $t_1=0.6 a/c$ and $t_2=3a/c$, 
and $P$ remains the same for $t>t_2$. For $x_1=2a$, $t_1=0.7 a/c$ and 
$t_2=2 a/c$, and $P$ returns to its initial profile at longer times. 
The coupling strength is $g=2.5 \sqrt{c}/a$. }
\end{figure}

As observables \eqref{Pi}, we choose smeared field operators 
$\Pi_\alpha=\int dx h(x-x_\alpha) \Pi(x)$ where $x_\alpha$ is a given 
position. We show in Appendix \ref{appsec} that the corresponding time 
functions \eqref{A}, \eqref{B} and \eqref{F} are
\begin{eqnarray}
A_\alpha^{(1)}(t) &=& (g/4) \left[ {\cal H}(x_\alpha-ct)
+{\cal H}(x_\alpha+ct)-2{\cal H}(x_\alpha) \right] , \nonumber \\
B_\alpha^{(1)} (t) &=& g \frac{ct}{2\pi} {\cal P} 
\int dx \frac{{\cal H}(x_\alpha+x)}{(ct)^2-x^2} , \nonumber \\
F_{12}(t) &=& \exp \left[ - \frac{2 g^2}{\pi c} \int dx 
\ln \left| 1+\frac{ct}{x} \right| {\cal H}(x) \right] , \label{1Dcase} 
\end{eqnarray} 
$A_\alpha^{(2)}=-A_\alpha^{(1)}$, and $B_\alpha^{(2)}=-B_\alpha^{(1)}$, 
where ${\cal H}(x)=\int dx' h(x') h(x-x')$ and ${\cal P}$ denotes the Cauchy 
principal value. For the Hamiltonian \eqref{H1D}, $F_{12}$ is real positive. 
Similar expressions are obtained for the field $\partial_x \phi$. The function 
$A_\alpha^{(\ell)}$ is nonvanishing essentially only for $x_\alpha$ close 
to $0$ where ${\cal S}$ is coupled to ${\cal M}$, and close to $\pm ct$. 
Classical correlations between the two systems propagate along ${\cal M}$ 
at velocity $c$. The time $|x_\alpha|/c$ appears also in the evolution of 
$B_\alpha^{(\ell)}$ which vanishes for $t$ close to this value provided 
$|x_\alpha| \gg a$. However, the behavior of this function is very different 
from that of $A_\alpha^{(\ell)}$ since it decays only as $t^{-1}$ at long times. 
The function $F_{12}$ vanishes algebraically in this limit. We remark that 
$F_{12}$ decays faster and faster as the temperature $T$ increases since 
$\ln F_{12}$ diverges with time as $Tt$ at finite $T$ \cite{QDS}. 

The coupling strength $g$ must be large enough to induce correlations 
between ${\cal S}$ and ${\cal M}$ but the larger $g$ is, the faster $F_{12}$ 
decreases with time, see \eqref{1Dcase}. As a consequence, practically 
only classical correlations between ${\cal S}$ and the observables 
$\Pi_\alpha$ can be observed, see Fig.\ref{fig:disp}. This figure shows 
the conditional probability distribution $P(\Pi_1=p | \sigma_x=1)$ of finding 
$\Pi_1=p$ immediately after a measurement of 
$\sigma_x=| 1 \rangle \langle 2 | + | 2 \rangle \langle 1 |$ with the result $1$, 
for ${\cal S}$ initially in the state 
$| u \rangle = 2^{-1/2}(| 1 \rangle+ | 2 \rangle)$. It reads 
$P=\big\langle | u \rangle\langle u | \otimes \delta(\Pi_1-p) \big\rangle
/\big\langle | u \rangle\langle u | \big\rangle$ where the numerator is given 
by \eqref{prob} with $u_{1/2}=2^{-1/2}$ and the denominator is equal to 
$[1+F_{12}(t)]/2$. The results in Fig.\ref{fig:disp} are obtained with 
the test function $h(x)=\exp(-x^2/a^2)$. For $x_1$ not too close to $0$, 
two time regimes can be distinguished.  In a short-time regime, $P$ is 
practically identical to the thermal Gaussian distribution determined by 
the the initial uncorrelated state \eqref{Omega}. For longer times, it is 
indistinguishable from that corresponding to the separable state 
\eqref{Oefft}, and shows (classical) correlations between ${\cal S}$ 
and ${\cal M}$ essentially for $t \simeq |x_1|/c$. The smaller $x_1$ is, 
the more noticeable the quantum interference part of \eqref{prob}, 
see Fig.\ref{fig:disp}. 

\subsection{Finite-range observables}\label{subsec:Fso}

The interaction-induced correlations between ${\cal S}$ and local degrees 
of freedom of ${\cal M}$ are then practically given by the separable 
state \eqref{Oefft}. On the other hand, we know that the quantum 
interference term of \eqref{Ot} is important at time $t=t_0$ for 
the observable \eqref{O12}. The corresponding field operator 
\eqref{Xi} can here be written in terms of $\Pi$ and $\phi$ as
\begin{equation}
\Xi (t_0) = g \left[ {\tilde \phi}(x_0) 
+ {\tilde \phi} (- x_0) - 2{\tilde \phi} (0) 
- \int_{-x_0}^{ x_0} dx {\tilde \Pi} (x)/c \right]  \label{Xi1D}
\end{equation}
where $x_0=ct_0$, ${\tilde \Pi} (x)=\int dx' h(x'-x) \Pi(x')$ and 
${\tilde \phi}$ is defined similarly, see Appendix \ref{appsec}. 
Thus, $\Xi (t_0)$ depends on a part of 
${\cal M}$ of extent essentially proportional to $t_0$. This suggests that, 
at any time, the difference between the actual state of ${\cal S}+{\cal M}$ 
and \eqref{Oefft} appears clearly if the physical fields $\Pi$ and 
$\partial_x \phi$ are measured in large enough regions. 

However, the observable \eqref{O12} is very particular. As a less peculiar 
example, let us consider the probability \eqref{prob} with $\Pi_1$ 
replaced by $\Pi_D=\int dx h_D(x) \Pi(x)$ where $h_D(x)$ is maximum 
at $x=0$ and vanishes for $|x| \gg D$. For this finite-range field operator, 
the time functions \eqref{A}, \eqref{B} and \eqref{F} are given by 
\eqref{1Dcase} with $\int dy h(y) h_D(x-y)$ in place of 
${\cal H}(x_\alpha+x)$. For $h(x)=\exp(-x^2/a^2)$ and 
$h_D(x)=\exp(-x^2/D^2)$, the corresponding factor ${\tilde F}_{12}$ 
in \eqref{prob}, satisfies, for $D \gg a$, 
\begin{multline}
{\bar g}^{-2} \ln {\tilde F}_{12} (t) \simeq -  \sqrt{\frac{2}{\pi}} \int dx 
\ln \left| 1+\frac{ct}{a} x^{-1} \right| e^{-x^2/2} \\ 
 +\frac{ 1 }{\pi} \left[ {\cal P} \int dx 
 \frac{\exp[-(ct/D)^2 x^2]}{1-x^2} \right]^2
\end{multline}
where $ {\bar g}=gac^{-1/2}$. Due to the presence of the above second 
term, ${\tilde F}_{12}$ decays more slowly than $F_{12}$. 
The characteristic time of this term is $D/c$ and hence it is significant 
at larger and larger times as the extent $D$ of $\Pi_D$ increases. 
However, since $D$ appears only via $ct/D$, this second term reaches 
its maximum at a time where it is far smaller than the first one. Therefore, 
even for large $D$, the difference between the actual state of 
${\cal S}+{\cal M}$ and \eqref{Oefft} cannot be revealed with the help of 
$\Pi_D$ for times larger than the decoherence time of ${\cal S}$ . 
This argumentation can be extended to arbitrary functions $h$ and $h_D$.     

\subsection{Possible relation with genuine disentanglement}\label{subsec:P}

We address here the following question : is the effective disentanglement 
found above simply a manifestation of genuine disentanglement ? 
As discussed in Section \ref{subsec:Iie}, the entanglement of ${\cal S}$ 
with ${\cal M}$ does not decrease with time. But that of ${\cal S}$ with 
a subsystem ${\cal S}'$ of ${\cal M}$ can. The rest of ${\cal M}$, named 
${\cal M}'$, constitutes the environment of ${\cal S}+{\cal S}'$ and may have 
the tendency to disentangle ${\cal S}$ and ${\cal S}'$ \cite{DH,JJ}. 
This environmental influence on ${\cal S}$ and ${\cal S}'$ competes with 
their mutual interaction that can be direct or mediated by ${\cal M}'$ 
\cite{RR,BFP}. Can the results obtained in the previous sections be 
explained by the dynamical behavior of the entanglement between 
${\cal S}$ and appropriate subsystems ${\cal S}'$ ?

To investigate this, we consider a portion ${\cal S}'$ specified by $|x|<D$ 
where $D$ is an arbitrary length. It can be shown, for large coupling 
strength $g$, that ${\cal S}$ and ${\cal S}'$ are entangled for $t<D/c$, as 
follows. We define the observables 
$A_{1/2}=\alpha \sigma_z\pm \beta \sigma_x$ where $\alpha^2+\beta^2=1$, 
$B_1=\sin(\gamma \Pi_0)$ where $\Pi_0=\int dx h(x) \Pi(x)$, and 
$B_2=\cos(\Xi (t_0)+\theta/2)$ where $\Xi (t_0)$ is given by \eqref{Xi1D} 
and $\theta/2$ is the phase of $\rho_{12}$. The eigenvalues of all 
these operators are in the interval $[-1,1]$. For $t_0 < D/c$ \cite{fn2}, 
$B_2$ is an observable of the system ${\cal S}'$ considered here. 
For $\gamma=\pi/4A_0^{(1)}(t_0)$ and $\alpha+i\beta=z/|z|$ where 
$z=\exp(-\gamma^2\langle \Pi_0^2 \rangle_{\cal M}/2)
+i |\rho_{12}| (1+F_{12}(t_0)^4 \cos \theta )$, we find 
\begin{equation}
\langle A_1(B_1 + B_2) + A_2(B_1 -B_2) \rangle (t_0) =2|z| ,
\label{BCHSH}
\end{equation}
see Appendix \ref{appsec:BCiv}. For non-entangled states, any average 
of this form satisfies the Bell-CHSH inequality \cite{B,CHSH}, i.e., is 
between $-2$ and $2$ \cite{W}. This is not the case here  for 
$g^2 \gg c/a^2$ since $\mathrm{Re} z \rightarrow 1$ in this limit.

Whereas ${\cal S}$ and ${\cal S}'$ are entangled at least until time $D/c$ 
where the extent $D$ of ${\cal S}'$ can be as large as we like, correlations 
of ${\cal S}$ with observables of ${\cal S}'$ are well described by 
the separable state \eqref{Oefft} for much shorter times. First, this is clear 
for the local field operators $\Pi_\alpha$ discussed in section 
\ref{subsec:Lo}. But this may simply mean that ${\cal S}$ and a small 
segment of ${\cal S}'$ located at $x=x_\alpha$ have disentangled. 
More interesting is the behavior of the operator $\Pi_D$ of the previous 
section. It is an observable of ${\cal S}'$ but not of any portion of 
${\cal S}'$. Thus, the corresponding effective disentanglement is not 
simply related to genuine disentanglement.

\section{Conclusion}

In summary, we have studied a measurement model in which 
the measured system ${\cal S}$ is linearly coupled to a measurement 
apparatus ${\cal M}$ that consists of harmonic oscillators. In general, 
the interaction between ${\cal S}$ and ${\cal M}$ entangles these 
two systems. This interaction-induced entanglement is important as 
it is the source of the decoherence of ${\cal S}$. However, we found 
that, though ${\cal S}$ and ${\cal M}$ get entangled with each other, 
correlations between ${\cal S}$ and physically relevant observables 
of ${\cal M}$ become classical with time. At long enough times, 
the corresponding expectation values are identical to that of a 
time-dependent classically correlated state which can be determined 
explicitly. Whereas this long-time state is the same for all 
the considered observables, it is a priori not the case for the decay 
time scale of the quantum contribution to correlations. For any given 
time, observables can be found for which the effective 
disentanglement process is not completed at this time but occurs 
later on.

In order to better understand this, we examined the special case 
of a two-level system ${\cal S}$ measured by 
a one-dimensional free field system ${\cal M}$. Our findings are 
the following. The interaction-induced correlations between ${\cal S}$ 
and local degrees of freedom of ${\cal M}$ are essentially classical. 
For such observables, the difference between the actual state of 
the complete system ${\cal S}+{\cal M}$ and the effective separable 
state mentioned above is noticeable only close to the point where 
${\cal M}$ is coupled to ${\cal S}$ and for times shorter than 
the decoherence time of ${\cal S}$. This difference can be 
evidenced at longer times with the help of finite-range observables 
but which are very specific combinations of field operators probably 
difficult to achieve in practice. We have also shown that the obtained 
decay of quantum correlations cannot be explained by a genuine 
disentanglement process between ${\cal S}$ and appropriate 
subsystems of ${\cal M}$.  

It would be of interest to examine whether such effective 
disentanglement exits for other physical observables and measuring 
devises. The question is also relevant to more general models 
describing both the decoherence and relaxation of an open system, 
or to large systems interacting with each other. It would be especially 
interesting to determine how general the spatiotemporal behavior of 
classical and quantum correlations obtained for the studied 
one-dimensional measurement apparatus is. 

\begin{appendix}

\section{Derivation of the generating function expression}\label{appsec:Dgf}

To evaluate the generating function \eqref{Kdef}, we first note that 
\begin{multline}
\prod_\alpha \exp( i X_\alpha \Pi_\alpha ) = 
\exp \Big( i \sum_\alpha X_\alpha \Pi_\alpha \Big) \\
\times \exp \Big[-i \sum_{\alpha<\alpha'} X_\alpha X_{\alpha'} \sum_q 
\mathrm{Im} \big(\mu_{\alpha' q} \mu_{\alpha q}^* \big) \Big] .
\end{multline}
Then, using the relation \eqref{U}, we write
\begin{multline}
e^{i t H_\ell} \exp \Big( i \sum_\alpha X_\alpha \Pi_\alpha \Big) 
e^{-itH_\ell}  \\
= \prod_q \exp \Big( i \sum_\alpha X_\alpha \big[ \mu_{\alpha q} 
a^{\dag}_{ \ell q}(t) + \mu_{\alpha q}^* a_{\ell q}(t) \big] \Big)
\label{Kdiagapp}
\end{multline}
where $a_{\ell q}(t)=\exp(-it \omega_q) a_q+\lambda_{\ell q}
[\exp(-it \omega_q)-1]/\omega_q$. 
Finally, with the thermal average
$\langle \exp(z a_q - z^* a^{\dag}_q ) \rangle_{\cal M}
=\exp[-|z|^2/2\tanh(\omega_q/2T)]$ and
\begin{equation}
\langle \Pi_\alpha \Pi_{\alpha'} \rangle_{\cal M} = \sum_q 
\frac{\mathrm{Re} \big(\mu_{\alpha' q} \mu_{\alpha q}^* \big)}
{\tanh(\omega_q/2T)}
+i\mathrm{Im} \big(\mu_{\alpha' q} \mu_{\alpha q}^* \big) ,
\end{equation}
we obtain the expression \eqref{Kdiag}. 

In the case $\ell \ne \ell'$, one has to evaluate the thermal average of 
the product of \eqref{Kdiagapp} by $\exp(itH_\ell)\exp(-itH_{\ell'})$. 
This factor can also be expressed as the exponential of a linear 
combination of the annihilation and creation operators $a_q$ and 
$a^{\dag}_q$. Doing so, we find \eqref{K} where the phase of 
$F_{\ell\ell'}=|F_{\ell\ell'}|\exp(i\varphi_{\ell\ell'})$ is 
\begin{multline}
\varphi_{\ell\ell'}=\sum_q \big( |\lambda_{\ell' q}|^2 
-|\lambda_{\ell q}|^2 \big)
\frac{\omega_q t-\sin(\omega_q t)}{\omega_q^2} \\
+4 \mathrm{Im} \big(\lambda_{\ell' q} \lambda_{\ell q}^* \big) 
\frac{\sin^2(\omega_q t)}{\omega_q^2} .
\end{multline}
  
\section{One-dimensional measurement apparatus}\label{appsec}

To derive the expressions \eqref{1Dcase}, we first consider a finite system 
${\cal M}$ described by the Hamiltonian 
\begin{equation}
H_0 = \frac{1}{2} \int_{-L}^L dx \left[ \Pi(x)^2 + c^2 (\partial_x \phi)^2 \right] 
= \sum_{q>0} cq \left( a^{\dag}_q a^{\phantom{\dag}}_q + \frac{1}{2} \right) 
\end{equation}
where $q=n\pi/2L$, $n \in \N$. In this case, the operators $\Pi(x)$ and 
$a_q$ are related by
\begin{equation}
\Pi(x) = \sqrt{\frac{c}{2L}} \sum_{q>0} \sqrt{q} \cos(qx+\theta_q)
 \left( a^{\dag}_q + a^{\phantom{\dag}}_q \right) \label{Piapp}
\end{equation}
where $\theta_q=0$ if $2Lq/\pi$ is even, and $\pi/2$ otherwise. 
Thus, for an even test function $h$, the coupling between ${\cal S}$ and 
${\cal M}$ given in \eqref{H1D} leads to 
$\lambda_{1q}=g(cq/2L)^{1/2} \int dx h(x) \cos(qx)=-\lambda_{2q}$ if 
$2Lq/\pi$ is even, and $0$ otherwise. 
For the smeared field operator $\Pi_\alpha=\int dx h(x-x_\alpha) \Pi(x)$, 
the coefficients $\mu_{\alpha q}$ are given by similar expressions. 
We remark that, since $\lambda_{2 q}=-\lambda_{1 q} \in \R$, 
$F_{12}=|F_{12}|$ here, see Appendix \ref{appsec:Dgf}. 

As $\lambda_{\ell q}=0$ when $2Lq/\pi$ is odd, the corresponding terms 
do not contribute to \eqref{A}, \eqref{B} and \eqref{F}. In the limit 
$L\rightarrow \infty$, the sums over the remaining $q$ become 
integrals. The functions \eqref{A} are, for example, given by 
\begin{multline}
A_\alpha^{(1)}(t) = \frac{g}{2\pi} \int_0^\infty dq \int dx 
h(x) \cos(qx) \\ \times \int dx' h(x'-x_\alpha) \cos(qx') [cos(cqt)-1]
\end{multline}
and $A_\alpha^{(2)}(t)=-A_\alpha^{(1)}(t)$. Similar expressions can 
be obtained for $B_\alpha^{(\ell)}(t)$ and $F_{12}(t)$ which finally 
give \eqref{1Dcase} after integration over $q$. 

The conjugate field to \eqref{Piapp} is
\begin{equation}
\phi(x) = \frac{i}{\sqrt{2cL}} \sum_{q>0} \frac{1}{\sqrt{q}} 
\cos(qx+\theta_q) \left(  a^{\phantom{\dag}}_q - a^{\dag}_q \right) .
\end{equation}
Using this expression and  \eqref{Piapp}, the observable \eqref{Xi} 
can be written as
\begin{multline}
\Xi (t_0) = g \int dx h(x) \int_{0}^{ x_0} dx' \partial_x \phi (x+x') 
- \partial_x \phi (x-x') \\ -\Pi (x+x')/c -\Pi (x-x')/c    
\end{multline}
where $x_0=ct_0$, which leads to \eqref{Xi1D}.

\section{Bell-CHSH inequality violation}\label{appsec:BCiv}

To obtain the Bell inequality violation discussed in section \ref{subsec:P}, 
we first define
\begin{equation}
f(t)=\langle A_1(B_1 + B_2) + A_2(B_1 -B_2) \rangle (t) 
\end{equation}
where $A_{1/2}=\alpha \sigma_z\pm \beta \sigma_x$, 
$B_1=\sin(\gamma \Pi_0)$ and $B_2=\cos(\Xi (t_0)+\theta/2)$. 
This function can be rewritten as
\begin{multline}
f(t)= 4 \beta \mathrm{Re} \rho_{21} 
\big\langle e^{i t H_1} B_2 e^{-i t H_2}  \big\rangle_{\cal M}  \\
+ 2\alpha \big[ \rho_{11} \big\langle e^{i t H_1} B_1 
e^{-i t H_1} \big\rangle_{\cal M}  - \rho_{22} \big\langle e^{i t H_2} B_1 
e^{-i t H_2} \big\rangle_{\cal M} \big] \label{f}
\end{multline}
with the help of \eqref{Ot}. The above last two expectation values can 
be evaluated using \eqref{Kdiag}. Since 
$A_\alpha^{(2)}(t)=-A_\alpha^{(1)}(t)$ for the one-dimensional 
system ${\cal M}$ considered in section \ref{sec:1DMA}, they are 
opposite of each other and hence $f$ does not depend on $\rho_{11}$ 
(and $\rho_{22}=1-\rho_{11}$). For $h(x)=\exp(-x^2/a^2)$, explicit 
expressions can be obtained with $\langle \Pi_0^2 \rangle_{\cal M}=c/2$ 
and $A_0^{(1)}(t)=(ga/2)(\pi/2)^{1/2}[\exp(-(ct/a)^2)-1]$.    

To evaluate the first term of \eqref{f}, we use 
$\exp[i\Xi (t_0)]=\exp(-iH_1t_0)\exp(iH_2 t_0)$. We find
\begin{multline}
f(t_0)= 2 \beta |\rho_{12}| \left( 1
+ \exp [-2\langle \Xi (t_0)^2 \rangle_{\cal M} ] \cos\theta \right) \\
+ 2\alpha e^{-\gamma^2 \langle \Pi_0^2 \rangle_{\cal M}/2} 
\sin \left[ 2\gamma A_0^{(1)}(t_0) \right]  
 \end{multline}
where $\theta/2$ is the phase of $\rho_{12}$. With this choice, 
the first term above is practically equal to $2 \beta |\rho_{12}|$ 
for times larger than the decoherence time of ${\cal S}$ as 
$\exp(-2\langle \Xi (t_0)^2 \rangle_{\cal M})=F_{12}(t_0)^4$. 
The value $f(t_0)$ is maximum as function of $\alpha$ and $\beta$, 
at $\alpha+i\beta=z/|z|$ where 
$z= \exp(-\gamma^2 \langle \Pi_0^2 \rangle_{\cal M}/2) 
\sin [ 2\gamma A_0^{(1)}(t_0) ] +i |\rho_{12}| 
( 1+ F_{12} (t_0)^4 \cos\theta )$. For large coupling strength $g$, 
the real part of $z$ is close to $1$ for $\gamma = \pi/4A_0^{(1)}(t_0)$. 
These values of $\alpha$, $\beta$ and $\gamma$ lead to 
\eqref{BCHSH}. We remark that for $\rho_{12}=0$, $|f(t)|<2$ as it 
must be since ${\cal S}$ and ${\cal M}$ are never entangled in 
this case. 

\end{appendix}


\begin{thebibliography}{99}
\bibitem{fn0} The corresponding entropy of entanglement is
${\cal E} (t) =-\sum_{\eta=\pm} 
[1+\eta F_{12}(t)]\log_2[(1+\eta F_{12}(t))/2]/2$.
\bibitem{fn1} In the special case $s=2$, $\ln |F_{\ell\ell'}|$ 
diverges only logarithmically and thus $F_{\ell\ell'} G$ does 
not necessarily vanish whereas ${\cal S}$ decoheres inevitably. 

At zero temperature, as $t\rightarrow \infty$, 
$\ln |F_{\ell\ell'}| \sim t^{1-s}$ for $s < 1$ 
and $\ln |F_{\ell\ell'}| \sim \ln t$ for $s = 1$.
\bibitem{fn2} Strictly speaking, the condition on $t_0$ depends 
on the test function $h$. We assume that $D \gg a$.
\end{thebibliography}
\end{document}